\documentclass[11pt,twoside]{./atmp}

\usepackage{amsmath,amssymb}
\usepackage{amsfonts}

\usepackage{epsf}
\usepackage{psfrag}
\usepackage{graphics}
\usepackage{mathrsfs} 

\usepackage[all]{xy}
\usepackage{color}

\def\be{\begin{equation}}
\def\ee{\end{equation}}
\def\bea{\begin{eqnarray}}
\def\eea{\end{eqnarray}}
\def\r{\rho}
\def\l{\lambda}
\def\eqq{\stackrel{\sigma}{=}}
\def\scri{{\mathscr{I}}}

\begin{document}

\title[Gravitational Collapse to Toroidal Black Holes]
{Gravitational Collapse to Toroidal and Higher Genus Asymptotically AdS Black Holes}


\author[Filipe C. Mena, Jos\'e Nat\'ario and Paul Tod]{Filipe C. Mena$^{\scriptsize 1,2}$, Jos\'e Nat\'ario$^{\scriptsize 2}$ and Paul Tod$^{\scriptsize 3}$}

\address{$^{\scriptsize 1}$Departamento de Matem\'atica \\ Universidade do Minho \\ 4710-057 Braga, Portugal}

\addressemail{fmena@math.uminho.pt}

\address{$^{\scriptsize 2}$Departamento de Matem\'atica\\ Instituto Superior T\'ecnico \\ 1049-001 Lisboa, Portugal}

\addressemail{jnatar@math.ist.utl.pt}

\address{$^{\scriptsize 3}$Mathematical Institute \\ University of Oxford \\ St Giles' 24-29 \\ Oxford OX1 3LB, U.K.}

\addressemail{tod@maths.ox.ac.uk}

\begin{abstract}
We match collapsing inhomogeneous as well as spatially homogeneous but anisotropic spacetimes to vacuum static exteriors 
with a negative cosmological constant and planar or hyperbolic symmetry. The collapsing interiors include the inhomogeneous solutions of Szekeres and of Barnes, which in turn include 
the Lema\^itre-Tolman and the McVittie solutions. The collapse can result in toroidal or higher genus asymptotically AdS black holes. 
\end{abstract}

\maketitle

\section{Introduction}

The process of gravitational collapse to a black hole in general relativity can be modelled by the matching of two appropriate exact solutions. For a zero cosmological constant $\Lambda$ there are several well-known examples in spherical symmetry, one of the simplest being the Oppenheimer-Snyder model which results from the matching of a Friedmann-Lema\^itre-Robertson-Walker (FLRW) dust metric to a Schwarzschild exterior. For $\Lambda\ne 0$
spherical cases the results may be less well-known: the matching of $\Lambda$-FLRW with Kottler has been done by Balbinot et al. \cite{ES-lambda}, Nakao \cite{Nakao} and Markovic \& Shapiro \cite{Shapiro}, and its $\Lambda$-Lema\^itre-Tolman counterpart by Lake \cite{Lake}. The matching of the dust $\Lambda$-Szekeres solution  with Kottler was recently studied by Debnath et al. \cite{Debnath} (following Bonnor \cite{Bonnor} for $\Lambda=0$), while the matching of a collapsing fluid with tangential pressure and $\Lambda\ne 0$ to Kottler has been investigated by Madhav et. al. \cite{Joshi}. The above examples give rise to spherical black hole solutions with $\Lambda\ne 0$.

The collapse to non-spherical black holes is less widely studied. Smith \& Mann \cite{Smith-Mann} have shown that one can match a collapsing $k=-1$ FLRW spacetime to an asymptotically anti-de Sitter (AdS) exterior, as a model of gravitational collapse to higher genus asymptotically AdS black holes. In a related work, Lemos \cite{Lemos} matched a flat FLRW metric to a radiating Vaidya exterior. Since their introduction by Lemos \cite{Lemos1}, the physics of toroidal and higher genus black holes has also been studied (see e.g. \cite{Brill, Vanzo, Lemos2, Lemos3}). Subsequently there has been considerable interest in toroidal and higher genus models of black holes, shells and horizons (see e.g. \cite{Horowitz}), partly due to the existence of a ``landscape'' of vacua states in string theory with $\Lambda$ positive, negative and zero (see e.g. \cite{Douglas}). 

In this paper we shall investigate generalisations of some of the above results by matching non-spherical inhomogeneous and anisotropic interiors (mostly taken from the excellent book of Krasi\'nski \cite{Krasinski}) to the non-spherical Kottler metric with $\Lambda<0$, thus describing a wider class of examples of inhomogeneous and anisotropic gravitational collapse to toroidal and higher genus black holes. 

The Kottler metric with $\Lambda<0$ has initial and final curvature singularities and a time-like null infinity, $\scri$. Any radial time-like geodesic has an endpoint at each singularity, crossing the future horizon of $\scri$ on its way to a maximum radius before falling back through the past horizon of $\scri$ to $r=0$ (see Figure~\ref{Kottler}). In a matching to a dust interior, such a geodesic is the outer boundary of the matter. The interior is part of a cosmological solution which in turn typically has initial and final singularities. As we shall see below, for the matching to FLRW, the initial and final singularities of the interior are matched to the initial and final singularities of the exterior, but this isn't necessarily true for all the matchings we find. However we may arrange that the initial singularities coincide and that the final singularity of the interior is behind the event horizon.

In what follows we use units such that the speed of light $c$ and Newton's gravitational constant $G$ satisfy $c=8\pi G=1$.

\section{Matching Theory in Brief}
\label{sec:match-conditions}
Let $(M^{\pm},g^{\pm})$ be spacetimes with non-null boundaries $\sigma^{\pm}$.
Matching them requires an identification of the boundaries, i.e. a pair of embeddings
$\Phi_\pm:\; \sigma \longrightarrow M^\pm$ with $\Phi_\pm(\sigma) = \sigma^{\pm}$,
where $\sigma$ is an abstract copy of either boundary. 
Let $\xi^i$ be a coordinate system on $\sigma$. Tangent vectors to
$\sigma^{\pm}$ are obtained by $f^{\pm \alpha}_i =and Higher Genus
\frac{\partial \Phi_\pm^\alpha}{\partial \xi^i}$, and we shall usually work with orthonormal combinations $e^{\pm \alpha}_i$ of the $f^{\pm \alpha}_i$. There are also  unique (up to 
orientation) unit normal vectors $n_{\pm}^{\alpha}$ to the boundaries. 
We choose them so that if $n_{+}^{\alpha}$ points into $M^{+}$ then
$n_{-}^{\alpha}$ points out of $M^{-}$ or viceversa. 
The first and second fundamental forms
are simply $q_{ij}^{\pm}= e^{\pm \alpha}_i e^{\pm \beta}_j
g_{\alpha\beta}|_{_{\sigma^\pm}},~
H_{ij}^{\pm}=-n^{\pm}_{\alpha} e^{\pm
\beta}_i\nabla^\pm_\beta e^{\pm \alpha}_j$. It is well know that the matching conditions (in the absence of shells) 
require the equality of the first and second fundamental forms on $\sigma^{\pm}$, i.e.
\begin{equation}
q_{ij}^{+}=q_{ij}^{-},~~~~~~
 H_{ij}^{+}=H_{ij}^{-}.
\label{eq:backmc}
\end{equation}
These will be referred to as {\em first} and {\em second matching conditions}, respectively.
If symmetries are present, one chooses the $e^{\pm \alpha}_i$  to reflect the symmetry and then $H^\pm_{ij}$ simplifies, as we shall see.
\\\\
It is a familiar fact that, in order to match a perfect fluid spacetime to a vacuum solution, the fluid's pressure has to vanish at the matching boundary. When the matching conditions are satisfied, this necessarily follows from the Hamiltonian constraint (on the time-like boundary). It is possible, but perhaps rather artificial, to match a perfect fluid interior with cosmological constant to a vacuum exterior with a different cosmological constant at a surface with non-zero pressure: the jump in the cosmological constant is chosen to match the jump in the pressure. However, we shall mostly be concerned with dust interiors with cosmological constant and vacuum exteriors with the same cosmological constant. The time-like boundary of the interior must be ruled by matter world lines, which for dust are geodesics, and the matching conditions require that the boundary is also ruled by time-like geodesics of the exterior. This is part of the condition (\ref{eq:backmc}) matching the second fundamental forms: if we take the tangent to the fluid world lines as $e^{-\alpha}_{1}$ then the geodesic condition implies the vanishing of $H^-_{11}$. If the interior is not dust, as for example with the inhomogeneous $p\ne 0$ perfect fluids of section \ref{Barnes}, the matter world-lines ruling the boundary are accelerating and now the accelerations must match. 
\section{Spacetimes and matching}

\subsection{The Kottler solution as the exterior}
To begin with, let us recall the Kottler solution which we shall always take as the exterior in what follows. The metric is
\be
ds^{2}_+=-VdT^2+V^{-1}dr^2+r^2(d\theta^2+\Sigma^2(\theta)d\varphi^2)
\label{K1}
\ee
with 
\be
V=b-\frac{2m}{r}-\frac{\Lambda}{3}r^2
\label{K2}
\ee
and $\Sigma(\theta)=\theta, \sin\theta, \sinh\theta$, according to $b=0,1,-1$, respectively. With $\Lambda<0$ and $m>0$, $V$ has a unique positive zero and this solution describes a black hole with planar, spherical or
hyperbolic symmetry respectively, on an asymptotically AdS background. For $b=0$ or $-1$ it is possible to make identifications of the 2-metric of constant $T$ and $r$ to obtain toroidal and higher genus black holes. Notice that $\scri$ will have the same topology (times $\mathbb{R}$).

\begin{figure}[h!]
\begin{center}
\psfrag{r=0 (past singularity)}{$r=0$ (past singularity)}
\psfrag{r=0 (future singularity)}{$r=0$ (future singularity)}
\psfrag{geodesic}{geodesic}
\psfrag{horizon}{horizon}
\psfrag{i+}{$i^+$}
\psfrag{i-}{$i^-$}
\psfrag{I}{$\mathscr{I}$}
\epsfxsize=.6\textwidth
\leavevmode
\epsfbox{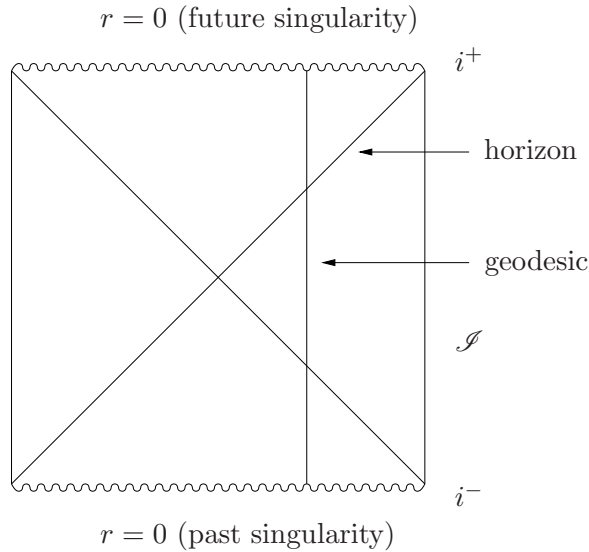}
\end{center}
\caption{Penrose diagram for the Kottler solution with $\Lambda < 0$, showing a typical
radial time-like geodesic at which the solution can be matched to an FLRW
interior. The horizon is the event horizon for the $\scri$ at the right.}\label{Kottler}
\end{figure}

Radial time-like geodesics of the metric (\ref{K1}) are characterised by the system
\begin{eqnarray*}
V\dot T&=&E\\
\dot r^2&=&E^2-V
\end{eqnarray*}
where $E$ is a constant which is strictly positive for a future-directed time-like geodesic which is ever outside the black hole. Therefore, with $\Lambda<0$, every such geodesic has a maximum value of $r$, and so has two end-points, past and future, at $r=0$ (cf.~Figure~\ref{Kottler}). For the dust interiors which follow, these geodesics are the candidate outer boundaries.

\hspace{1cm}\\\\
{\bf (a) Toroidal Kottler:}
\\\\
In the case $b=0$, we introduce Cartesian coordinates $(y,z)$ in place of $(\theta,\phi)$. To perform the matching preserving the symmetry we shall take a 3-surface $\sigma^+$ in the exterior generated by the orthonormal tangent vectors
\[e^+_1=\dot T\partial_T+\dot r\partial_r,~~~~e_2^+=r^{-1}\partial_y,~~~~e_3^+=r^{-1}\partial_z
\]
with $e^+_1$ tangent to geodesics, and parametrized by
\[
\Phi^+(\l,\tilde y,\tilde z)= (T(\l), r(\l), \tilde y, \tilde z).
\] 
An overdot will mean differentiation with respect to $\l$, which is proper time along the geodesics ruling the boundary. Thus in particular
\be
V\dot T^2-V^{-1}\dot r^2=1,
\label{norm}
\ee
which, together with the constancy of $\dot TV$, is equivalent to the geodesic equation. The induced metric on $\sigma^+$ is
\be
ds^{2}_+|_\sigma = -d\lambda^2+r^2(d\tilde y^2+d\tilde z^2).
\label{m1}
\ee
In all cases below, the normal from the exterior will be taken to be
\begin{equation}
\label{normal-kottler}
n^+=\frac{\dot r}{V}\partial_T+V\dot T\partial_r
\end{equation}
and the only non-zero terms of the second fundamental form are
\begin{equation}
\label{H-kottler}
H^+_{22}=H^+_{33}=\frac{V\dot T}{ r}.
\end{equation} 
\\\\
{\bf (b) Higher Genus Kottler:}
\\\\
With $b=-1$, we can take the hyperbolic metric on the surfaces of constant $(T,r)$ in the upper-half-plane form, so that the spacetime metric is
\[
ds^{2}_+=-Vdt^2+V^{-1}dr^2+\frac{r^2}{y^2}(dy^2+dz^2).
\]
The induced metric on the matching surface is
\be
ds^{2}_+|_\sigma = -d\lambda^2+\frac{r^2}{\tilde y^2}(d\tilde y^2+d\tilde z^2).
\label{met3}
\ee
The normal to the matching surface can be written as in (\ref{normal-kottler}) and the non-zero components of the second fundamental in this case 
are again given by (\ref{H-kottler}).
\\\\
In the next sections, we shall describe the different families of collapsing interior spacetimes to be considered, as well as their matching to the Kottler exterior.
We shall divide our analysis into two parts, corresponding to spatially homogeneous and inhomogeneous collapsing spacetimes.
\subsection{Spatially homogeneous collapsing spacetimes as interiors}
\subsubsection{FLRW spacetimes}
It is convenient to write the FLRW solutions as:
\begin{equation}
\label{FLRW}
ds^{2}_-=-dt^2+R^2(t)(d\rho^2+f^2(\rho)(d\theta^2+g^2(\theta)d\varphi^2)),
\end{equation}
with the functions $f$ and $g$ as given in Table \ref{table-FLRW}, for the different values of the curvature $k$.
\begin{table}
\begin{tabular}{|l|l|l|}
\hline
$k=1$ & $k=0$ & $k=-1$\\
\hline
$f=\sin\r,\cos\r;$ & (a) $f=1=g$ & (a) $f=e^{\pm \rho};\, g=1$\\
$g=\sin\theta,\cos\theta$ & (b) $f=\rho;\, g=\sin\theta, \cos\theta$ & (b) $f=\sinh\rho;\, g=\sin\theta, \cos\theta$\\
&&(c) $f=\cosh\rho;\, g=\sinh\theta, \cosh\theta$\\
\hline
\end{tabular}
\caption{The $f$ and $g$ functions of FLRW metric (\ref{FLRW}) depending on the curvature $k$.}
\label{table-FLRW} \end{table}
The cases with $k=1$ all correspond to spherical symmetry. For $k=0$ or $-1$, cases (a) have planar symmetry, (b) have spherical symmetry and (c) have hyperbolic symmetry.

As we noted in the introduction, the matching of a spherical form of the FLRW metric to a spherical exterior is familiar. The possible matchings of the non-spherical forms of FLRW metrics as interior to a non-spherical exterior Kottler solution can be summarized as follows:
\\\\
(i) $k=0$ and $b=0$: Cannot match. This fact was given in \cite{Smith-Mann} and \cite{Lemos}. In the latter reference, this FLRW was matched to a $b=0$ Vaidya exterior instead.\\\\
(ii) $k=-1$ and $b=-1$: Matched in \cite{Smith-Mann}.
\\\\(iii) $k=-1$ and $b=0$: Can match. 
\\\\
The third possibility was mentioned but not carried out in \cite{Smith-Mann}. To see how it works, take the FLRW from the third column of Table \ref{table-FLRW} with $f=e^\r$ and $g=1$ and take $\sigma^-$ to be $\r=\r_0$. The metric on the boundary is
\be
ds^2_-|_\sigma=-dt^2+R^2(t)e^{2\r_0}(d\theta^2+d\varphi^2),
\label{m2}
\ee
which must match to (\ref{m1}), so that $(t,\theta,\phi)=(\l,\tilde y,\tilde z)$ and %
\be
r\eqq R(t)e^{\r_0},
\label{mat1}
\ee
where we introduce the notation $\eqq$ for an equality holding at the matching surface. The second fundamental form from the interior has nonzero components just
\[H^-_{22}=H^-_{33}=\frac{1}{R(t)},\]
which must match (\ref{H-kottler}), so that
\be
V \dot T\eqq e^{\r_0}.
\label{mat2}
\ee
The matching will be possible if the proper time dependence imposed on $r$ by (\ref{mat1}) is consistent with it defining a geodesic. Since (\ref{mat2}) holds, this is equivalent to the normalisation condition (\ref{norm}), which, with the aid of (\ref{mat1}) and (\ref{mat2}), becomes
\be
\dot R^2=1+\frac{2me^{-3\r_0}}{R}+\frac{\Lambda}{3}R^2.
\label{mat2a}
\ee
This can be recognised as the Friedmann equation for the interior, with a dust density $\mu$ satisfying
\be
\frac16 R^3\mu=me^{-3\r_0}.
\label{mat3}
\ee
Thus (\ref{mat3}) gives the mass $m$ of the exterior in terms of the matter in the interior. Every solution of (\ref{mat2a}) that expands from an initial singularity has a maximum value of $R$ and then recollapses, just as the time-like geodesics in the exterior do. By (\ref{mat1}), the matching runs all the way from bang to crunch. Note that, even with the boundary identified as a torus, the interior is not spatially compact, as the range of $\r$ goes to $-\infty$.

\subsubsection{Anisotropic spacetimes as interiors} \label{sectionFarnsworth}
 As an example of a homogeneous anisotropic interior, we consider the tilted dust LRS Bianchi V metric given by the Farnsworth solution \cite{Farnsworth,Kramer}, which reduces to case (iii) of the previous section when the tilt vanishes. The metric is:
\be
ds^{2}_-=-dt^2+f^2d\r^2+g^2 e^{2\r}(dy^2+dz^2).
\label{F1}
\ee
Here $f$ and $g$ are functions of a single variable $u=t+\alpha\r$, $a$ and $\alpha$ are constants and
\be
f=a(\alpha g_u+g),
\label{f1}
\ee
while $g$ satisfies
\[
2gg_{uu}+ g_u^{\;2}-\Lambda g^2-\frac{1}{a^2}=0,
\]
which has the first integral
\be
gg_u^{\;2}=\frac{\Lambda}{3}g^3+\frac{g}{a^2}+c
\label{fe1}
\ee
with $c$ another constant (compare (\ref{fe1}) with (\ref{mat2a})). The fluid flows along the vector field $u=\partial/\partial t$ but the surfaces of constant $t$ are not the surfaces of homogeneity: this is what it means for the fluid to be tilted. The energy-density is given by
\[
\mu=\frac{6(\frac{1}{3}\Lambda g-g_{uu})}{\alpha g_u+g}
\]
which, with the aid of (\ref{fe1}), simplifies to
\be
\mu=\frac{3ac}{fg^2}.
\label{mu1}
\ee
The signs of the constants need to be chosen so that $f$ and $\mu$ are positive. As we shall see below, we need $c$ positive, so also $a>0$. Solutions of (\ref{fe1}) expand from $g=0$ at say $u=0$, to a maximum value $g_{max}$ and then contract back to $g=0$ at say $u=u_{F}$. Near the beginning $g\sim u^{2/3}$ while near the end $g\sim (u_F-u)^{2/3}$. From (\ref{f1}), if we choose $\alpha>0$, then $f\sim u^{-1/3}$ near $u=0$, with a positive constant of proportionality, but then $f$ would be negative near $u=u_F$, and so must vanish before $u=u_F$. This will be a curvature singularity (since $\mu$ will diverge). It must happen in the contracting phase (so that $g_u<0$) and will occur when $g$ is the unique positive root of the cubic
\be
G(g):=\left(\frac{\Lambda}{3}-\frac{1}{\alpha^2}\right)g^3+\frac{g}{a^2}+c,
\label{cub1}
\ee
obtained by equating $f$ to zero and using (\ref{fe1}). It will be important, once we have matched this interior to the planar Kottler exterior, that this singularity lies inside the event horizon.

The Farnsworth solution generalises the $k=-1$ FLRW solution which can be recovered for $a=1$ and $\alpha=0$.
We shall now show that this solution can be matched to Kottler with $b=0$.
The matching surface $\sigma^-$ is parametrized by
\[
\Phi^-(\l, \tilde y, \tilde z) = (\l, \r_0, \tilde y, \tilde z).
\] 
The first fundamental form in this case is
\[
ds^{2-}|_{\sigma}=-d\l^2+g^2e^{2\r_0}(d\tilde y^2+d\tilde z^2)
\]
so the first matching condition is
\be
r\eqq ge^{\r_0}.
\label{mat4}
\ee
The unit normal to the matching surface is
\[
n^-=\frac{1}{f}\partial_\r
\]
and the second fundamental form is found to have non-zero components
\[
H^-_{22}=H^-_{33}=\frac{1}{f}\left(\alpha\frac{g_u}{g}+1\right)
\]
so that the second matching condition is 
\[
\frac{V\dot T}{r}\eqq \frac{1}{f}\left(\alpha\frac{g_u}{g}+1\right)
\]
which, with (\ref{mat4}), implies 
\[
V\dot T\eqq \frac{e^{\r_0}}{a},
\]
a constant, as required. For (\ref{norm}) to hold, we need
\[
g_u^{\;2}=\frac{1}{a^2}+\frac{\Lambda}{3}g^2+\frac{2me^{-3\r_0}}{g}
\]
which is the remaining field equation (\ref{fe1}) if we make the identification
\[
c\eqq 2me^{-3\r_0},
\]
so that indeed $c>0$.

\begin{figure}[b!]
\begin{center}
\psfrag{past singularity}{past singularity}
\psfrag{r=0 (Kottler future singularity)}{$r=0$ (Kottler future singularity)}
\psfrag{f=0 (Farnsworth future singularity)}{$f=0$ (Farnsworth future singularity)}
\psfrag{matching surface}{matching surface}
\psfrag{apparent horizon}{apparent horizon}
\psfrag{horizon}{horizon}
\psfrag{Cauchy horizon}{Cauchy horizon}
\psfrag{i+}{$i^+$}
\psfrag{i-}{$i^-$}
\psfrag{I}{$\mathscr{I}$}
\psfrag{r=-i}{$\rho=-\infty$}
\epsfxsize=1.0\textwidth
\leavevmode
\epsfbox{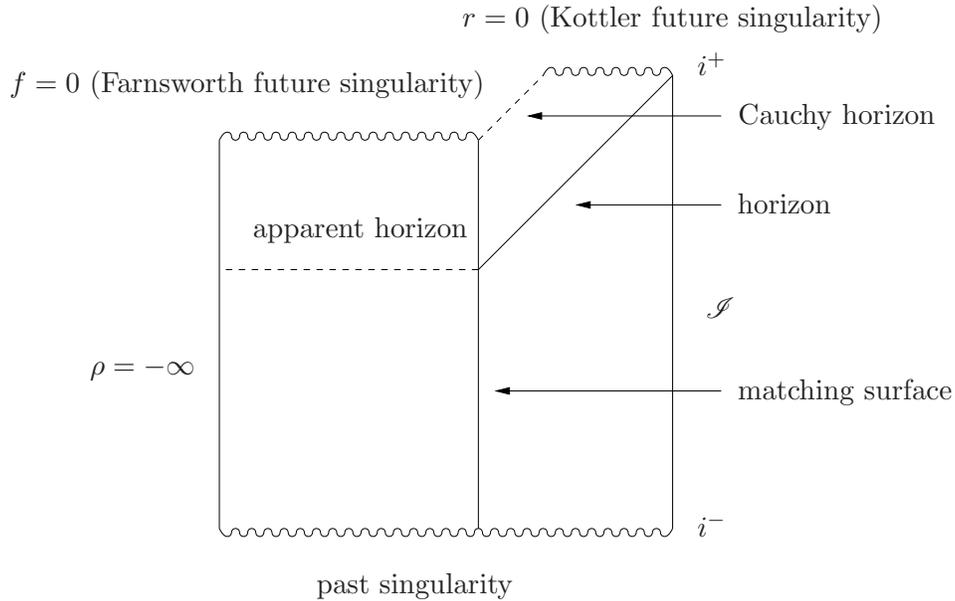}
\end{center}
\caption{Penrose diagram for the matching of Farnsworth to Kottler with $\Lambda<0$. The matchings for Szekeres interiors are qualitatively similar.}\label{Farnsworth}
\end{figure}

This completes the matching, which will hold from the initial singularity until $f$ vanishes. Then it fails and the exterior can no longer be the Kottler solution. However, the vanishing of $f$ can happen inside the event horizon, when the exterior solution will be Kottler up to and beyond the event horizon (see Figure~\ref{Farnsworth}).

The vanishing of $f$ happens when $g$ is a root of (\ref{cub1}). This will be inside the horizon if $G(g_H)<0$, where $g_H$ is the value of $g$ corresponding to the location of the horizon, $r=r_H = \left(-\frac{6m}{\Lambda}\right)^{1/3}$. The condition is
\be
\alpha<ag_H=a\left(-\frac{3c}{\Lambda}\right)^{1/3},
\label{cond1}
\ee
which is therefore a condition on the parameters of the interior solution. It is easy to check that this condition delays the final singularity in the interior until after the formation of a marginally-outer-trapped surface (which occurs when $g_u=-a^{-1}$, or, using (\ref{fe1}), when $g=g_H$).

\subsection{Inhomogeneous collapsing spacetimes as interiors}
\subsubsection{A higher genus analogue of the Farnsworth solution}\label{sectionhigherFarnsworth}

A dust solution (in the Szekeres family) which is a higher genus analogue of the Farnsworth solution, but not itself homogeneous, may be constructed as follows. The metric (compare (\ref{F1})) is
\be
ds^2=-dt^2+f^2d\rho^2+g^2\cosh^2\rho\frac{(dy^2+dz^2)}{y^2}
\label{F3}
\ee
where $f$ and $g$ are functions of a variable $u=t+\alpha\log\cosh\rho$ for constant $\alpha$. Then the Einstein equations are satisfied for dust with 4-velocity $\partial/\partial t$ and cosmological constant $\Lambda$ if
\[
f=\alpha g_u+g,\]
and $g$ satisfies
\[
gg_u^{\;2}=\frac{\Lambda}{3}g^3+g+c,
\]
in terms of another (positive) constant $c$. The dust density is 
\[
\mu=\frac{3c}{fg^2}
\]

This solution is not homogeneous on the surfaces of constant $u$. It can be matched to the hyperbolic Kottler solution at $\r=\r_0$ if 
\begin{eqnarray*}
r&\eqq & g\cosh\r_0\\
V \dot T&\eqq & \sinh\r_0
\end{eqnarray*}
when the normalisation condition (\ref{norm}) reduces to
\[c\eqq 2m\cosh^{-3}\r_0.\]
Just as in the previous case, with a positive $\alpha$ there exist solutions which expand from a singularity at which $g$ vanishes and $f$ diverges, pass through a maximum $g$, and recontract. In the contracting phase, $f$ will vanish before $g$ and this will mark the final singularity. For this to lie inside the future horizon there is a condition analogous to (\ref{cond1}), which this time is
\[\alpha<g_H\coth\r_0.\]
The interpretation is as before: this condition ensures that a marginally-outer-trapped surface forms on the boundary before the interior becomes singular. A difference from the previous case is that $\r_0$ appears explicitly in the condition, and now, given the parameters $(c,\alpha,\Lambda)$ determining the interior, there will always be a (small enough) $\r_0$ at which to make the matching.

\subsubsection{The Szekeres solutions}
We next consider the Szekeres form of the metric 
\[
ds^2=-dt^2+e^{2\alpha}d\r^2+e^{2\beta}(dy^2+dz^2)
\]
where $\alpha$ and $\beta$ are two $C^2$ arbitrary functions of $t,\r,y,z$. This is compatible with a perfect fluid source with 4-velocity $u^i=\delta^i_0$ and pressure $p=p(t)$. We only consider the case where $\partial \beta/\partial \r\ne 0$, which is the Szekeres class that generalises the Lema\^itre-Tolman spherically-symmetric dust solution. This is because we need the 2-surface of constant $(t,\r)$ to have non-zero second fundamental form in the 3-surface of constant $t$. 

The Einstein field equations for a perfect fluid imply
\begin{eqnarray}
e^\beta&=& \phi(t,\r) e^{\nu}\nonumber\\
e^\alpha& =& h(\r) e^{-\nu} \partial_\r e^\beta\label{alp}\\
e^{-\nu}&=& A(\r) (y^2+z^2)+2 B_1(\r) y+2B_2(\r) z+C(\r)\label{nu}
\end{eqnarray} 
where $A, B_1, B_2, C$ and $h$ are arbitrary $C^2$ functions of $\r$. The 2-surfaces of constant $(t,\r)$ have a metric which is conformal to
\[e^{2\nu}(dy^2+dz^2),\]
with a conformal factor independent of $y$ and $z$. From the form (\ref{nu}) of $\nu$, this 2-surface has constant positive, zero or negative curvature according as $AC-B_1^2-B_2^2$ is positive, zero or negative and we can conveniently call these the spherical, flat or hyperbolic cases respectively.

To obtain a dust source with cosmological constant, we set $p(t)=-\Lambda$ and then from \cite{BSS} or \cite{Krasinski} we find
\begin{eqnarray}
\dot \phi^2& =& -k(\r)+\frac{2M(\r)}{\phi}+\frac{1}{3}\Lambda\phi^2\nonumber\\
\mu& =& \frac{\partial_\r(2Me^{3\nu})}{e^{2\beta}\partial_\r(e^\beta)}\nonumber
\end{eqnarray}
with 
\be
AC-B_1^2-B_2^2=\frac{1}{4}(h^{-2}(\r)+k(\r)).
\label{norm2}
\ee
Note the scaling freedom:
\begin{eqnarray*}
(\hat A, \hat B_i, \hat C)&=&e^{\zeta(\r)}(A,B_i,C)\\
\hat\nu&=&\nu-\zeta\\
\hat h&=&e^{-\zeta}h\\
\hat k&=&e^{2\zeta}k\\
\hat\phi&=&e^{\zeta}\phi
\end{eqnarray*}
and the freedom to perform affine transformations and inversions in $y$ and $z$.

The spherical case has already been matched to the spherical Kottler solution \cite{Lake}, so here we shall perform the matching of the flat (including toroidal) and hyperbolic (including higher genus) Szekeres solutions to the corresponding Kottler solutions. We shall match
across $\r=\r_0$, and use affine transformations in $y$ and $z$ to simplify $A$, $B_i$ and $C$ at $\r_0$. 

It is worth noting that the matchings we obtain for the non-spherical Szekeres solutions specialise to give matchings of the non-spherical Lema\^itre-Tolman solutions (for these solutions see \cite{Krasinski}). It is also worth noting that the Farnsworth solution of section \ref{sectionFarnsworth} and its analogue of section \ref{sectionhigherFarnsworth} fall into the Szekeres class of solutions, respectively planar and hyperbolic. Consequently one should expect that, as in those examples, there will be restrictions on the range of time over which the matchings that we find will hold, arising as conditions on the interior.
\\\\
{\bf (a) Toroidal Szekeres:}
\\\\
Without loss of generality, we use the freedoms noted above to choose $A=B_i=0$, $C=1$ at $\r_0$. The first fundamental form on the matching surface can be written as
\be
ds^{2}_-|_\sigma = -d\lambda^2+\phi^2(d\tilde y^2+d\tilde z^2)\label{sz1}
\ee
so that the first matching condition gives
\be
\phi (\l, \r)\eqq r.
\label{sz1a}
\ee
The normal to the matching surface is
\[
n^- = e^{-\alpha}\partial_\r,
\]
and the second fundamental form at $\sigma$ has non-zero components
\begin{eqnarray}
H^-_{22}=H^-_{33}&=& \frac{1}{\phi h}\label{sz2}
\end{eqnarray} 
so that the second matching condition is
\begin{eqnarray}
V\dot T& \eqq& \frac{1}{h}.
\end{eqnarray}
From (\ref{norm}) we obtain the matching of the mass:
\[M(\r_0)=m.\]
This completes the matching. With $M(\r)$ and $M_{,\r}(\r)$ both positive (as is physically natural), the density is singular where either of $\phi$ or $\beta_{,\r}$ vanishes. Vanishing $\phi$ is matched, via (\ref{sz1a}), to the singularity in the exterior, but vanishing $\beta_{,\r}$ entails vanishing of $e^{\alpha}$ via (\ref{alp}). This is like the problem of vanishing $f$ in the previous two examples but, because of the extra functional freedom in the general Szekeres solution, we cannot write down an explicit condition like (\ref{cond1}) to force this to happen behind the horizon. Rather we simply insist, as a condition on the interior, that $\beta_{,\r}$ be positive for values of $\phi$ giving values of $r$ outside the horizon.
\\\\
{\bf (b) Higher genus Szekeres:}
\\\\
By suitable transformations of $(y,z)$ and scaling, we may assume without loss of generality that $A=B_2=C=0$, $B_1=1/2$ at $\r_0$. The induced metric on the boundary can be written as
\[
ds^{2}_-|_\sigma=-d\l^2+\frac{\phi^2}{\tilde y^2} (d\tilde y^2+d\tilde z^2)
\] 
and the first matching condition is
\be
\phi\eqq r.
\ee
The unit normal to $\sigma$ is
\[
n^- = e^{-\alpha} \partial_\r
\]
so that the second matching condition is
\[
V\dot T\eqq \frac{1}{h}.
\]
The normalisation condition (\ref{norm}) reduces to 
\begin{eqnarray*}
 M(\r_0) = m
\end{eqnarray*}
which is continuity of the mass again.

The comments at the end of the previous section regarding the range over which the matching holds apply in this case also.


\subsubsection{Barnes solutions}
\label{Barnes}
The Barnes solutions \cite{bar1} can be written in the form \cite{Krasinski}
\[
ds^2=-D^2dt^2+W^{-2}(dx^2+dy^2+dz^2)
\]
where $W=W(t,x,y,z)$ is determined, in a way which varies from case to case, by another function $w=w(t,\r)$, with $\r=\r(x,y,z)$, satisfying
\be
\frac{w_{,\r\r}}{w}=f(\r)
\label{ee2}
\ee
and 
\[
D=F\frac{W_{,t}}{W}
\]
with $F=F(t)$ an arbitrary function. 
These solutions, which include the well-known McVittie class, may be characterised as the general solution of the Einstein equations with the perfect-fluid form of stress-tensor, with a twist-free, shear-free but expanding 4-velocity. In general they don't have a barotropic equation of state (see the account in \cite{Krasinski}). A solution is determined by a choice of $F(t)$ and $f(\r)$, with data $w(t,\r_0)$ and $w_{\r}(t,\r_0)$ for (\ref{ee2}) at some value $\r_0$ of $\r$.

We require as before that the boundary be ruled by matter flow-lines, and that the pressure vanishes at the boundary, but since these are not dust solutions, we don't require that the matter flow-lines ruling the boundary be geodesics.

Some examples of matching in the spherical case have been done already (see e.g \cite{cahmcv} and other references in \cite{Krasinski}) so we shall again consider only the plane and hyperbolic cases. 
\\\\
{\bf (a) Toroidal Barnes:} 
\\\\
These solutions have
\[
W=w(t,\r),~~~\text{with}~~\r=x
\]
and
\begin{eqnarray}
\mu+\Lambda&=&\frac{3}{F^2}+2fw^2-3w^2_{,\r}\\
\label{mu-barnes-1}
p-\Lambda&=& -\frac{3}{F^2}+w^2_{,\r}-2\frac{F_{,t}}{F^2D}-2\frac{Fw_{,\r}}{Dw}(ww_{,t\r}-w_{,t}w_{,\r}).\label{p-barnes-1}
\end{eqnarray}
We take the matching surface 
from the exterior to be given in terms of proper time $\l$ by $(T(\l), r(\l))$ and from the interior by $(t(\l), \r=\r_0)$. 
The first matching conditions are
\be
V\dot T^2-V^{-1}\dot r^2\eqq D^2\dot t^2\eqq 1\label{mat5}\ee
and
\be
r\eqq \frac{1}{w}\label{mat6}
\ee 
From (\ref{mat6}) we obtain
\be
\dot r=-\frac{1}{Fw}.
\label{mat11}
\ee
The normal to the matching surface is
\[
 n^-= w\partial_\r
\]
and the second fundamental form gives
\begin{eqnarray*}
H^-_{11}&=& -w\frac{D_{,\r}}{D}\\
H^-_{22}&=&H^-_{33}=-Dw_{,\r}\dot t.
\end{eqnarray*} 
The second matching conditions are
\begin{eqnarray}
-w_{\r}&\eqq&\frac{V\dot T}{r}\label{mat7}\\
-w\frac{D_{,\r}}{D}&\eqq& \ddot T\dot r-\ddot r\dot T-\frac{1}{2}VV_{,r}\dot T^3+\frac{3}{2}V^{-1}V_{,r}\dot T\dot r^2\label{mat8}
\end{eqnarray}
where the extra complexity of (\ref{mat8}) reflects the fact that the boundary is not ruled by geodesics; the right-hand-side in (\ref{mat8}) is (minus) the inner product of the acceleration in the exterior of the boundary curve with the unit normal.

The normalisation condition (\ref{norm}), with (\ref{mat11}) and (\ref{mat7}), leads to
\be
V\eqq\frac{w^{2}_{,\r}}{w^2}-\frac{1}{F^2w^2}
\label{mat9}
\ee
and then preservation of this in time implies, by (\ref{p-barnes-1}),
\be
p\eqq 0
\label{mat10}
\ee
which also results from (\ref{mat8}), given the rest of the matching. Thus, as expected, the pressure must vanish at the matching surface.
\\\\
One may ask, given a planar Barnes solution, where should the matching surface be? We need a (necessarily comoving) surface of constant $\r$, say $\r=\r_0$, at which $p=0$. Then the quantity $Q$, defined by
\[
Q=\frac{1}{2w^3}\left(\frac{1}{F^2}-w_{\r}^2-\frac{\Lambda}{3}\right),
\]
is necessarily constant on $\r=\r_0$, by virtue of the field equations. We match to an exterior planar Kottler solution according to (\ref{mat5}) and (\ref{mat6}) and then we satisfy (\ref{mat9}) provided the parameter $m$ of the Kottler exterior equals $Q$. 

Not every Barnes solution will contain a constant-$\r$ surface at which $p$ vanishes. However it is possible to construct examples by using the matching surface as the surface at which data for (\ref{ee2}) is given. That is, we choose $F(t)$ and $f(\r)$, and, then given a choice of boundary $\r=\r_0$, we choose $w(t,\r_0)$ (positive) and $w_{\r}(t,\r_0)$ so that the density $\mu(t,\r_0)$ from (\ref{mu-barnes-1}) is positive, the pressure $p(t,\r_0)$ from (\ref{p-barnes-1}) is zero, and the quantity $Q(\r_0)$, which is now necessarily constant in time, is positive (we also want $D$ positive and, to be matching outside the black hole, $w_{,\r}$ negative). Integration of (\ref{ee2}) then gives a Barnes solution, at least in a neighbourhood of the boundary, which has a boundary with the desired properties. In the exterior, the boundary follows a curve given by (\ref{mat6}) together with the integration of (\ref{mat7}), using $\dot t\eqq D^{-1}$.
\\\\
{\bf (b) Higher genus Barnes:}
\\\\
In this case 
\[
W=y w(t,\r)~~~\text{with}~~\r=x/y
\]
and therefore the metric can be written as
\[
ds^2=-D^2dt^2+\frac{1}{y^2w^2}(y^2d\r^2+2\r yd\r dy+(1+\r^2)dy^2+dz^2)
\] 
where we have changed variables to eliminate $x$, together with
\begin{eqnarray}
\mu+\Lambda&=&\frac{3}{F^2}+2(\r^2+1)fw^3+6\r ww_{,\r}-3(\r^2+1)w_{,\r}^2-3w^2 \label{mu-barnes-2}\\
p-\Lambda&=&-\frac{3}{F^2}-2\r ww_{,\r}+(\r^2+1)w^2_{,\r}+w^2-2\frac{F_{,t}}{F^2D}\nonumber\\
&&+2\frac{F}{D}\left(\r-(\r^2+1)\frac{w_{,\r}}{w}\right)(ww_{,t\r}-w_{,t}w_{,\r})\label{p-barnes-2}
\end{eqnarray}
We take the matching surface 
from the exterior to be given by $(T(\l), r(\l))$ and from the interior $(t(\l), \r=\r_0)$. 
The first matching conditions are
\be
V\dot T^2-V^{-1}\dot r^2\eqq D^2\dot t^2=1\label{hb1}
\ee
and
\be
r^2\eqq \frac{1+\r^2}{w^2}
\label{hb2}
\ee
The normal to the matching surface is
\[
 n^-= w\sqrt{1+\r^2}\partial_\r-\frac{w\r y}{\sqrt{1+\r^2}}\partial_y
\]
and the second fundamental form is
\begin{eqnarray*}
H^-_{11}&=& -w\sqrt{1+\r^2}\frac{D_{,\r}}{D}\\
H^-_{22}&=&H^-_{33}=\frac{w\r}{\sqrt{1+\r^2}}-\sqrt{1+\r^2}w_{,\r}
\end{eqnarray*} 
so that the second matching conditions are
\begin{eqnarray}
\r-(1+\r^2)\frac{w_{,\r}}{w}&\eqq&V\dot T\label{hb3}\\
-w\sqrt{1+\r^2}\frac{D_{,\r}}{D}&\eqq& \ddot T\dot r-\ddot r\dot T-\frac{1}{2}VV_{,r}\dot T^3+\frac{3}{2}V^{-1}V_{,r}\dot T\dot r^2.\label{hb4}
\end{eqnarray}
It follows from (\ref{hb1}) and (\ref{hb3}) that
\[V\eqq \left(\r-(1+\r^2)\frac{w_{,\r}}{w}\right)^2-(1+\r^2)\frac{1}{F^2w^2},
\]
and then it is straightforward to confirm that
\[
p\eqq 0.
\]
One argues as before for the existence of higher-genus Barnes solutions with a constant-$\r$ surface at which $p$ vanishes: choose the free functions $F(t)$, $f(\r)$ and the data $w(t,\r_0)$, $w_{,\r}(t,\r_0)$ so that $p(t,\r_0)=0$; propagate inwards with (\ref{ee2}) to obtain the interior; the quantity
\[Q=\frac{(\r^2+1)^{1/2}}{2w^3}\left(-w^2-\frac{\Lambda(\r^2+1)}{3}+\frac{(1+\r^2)}{F^2}-(\r w-(\r^2+1)w_{,\r})^2\right)
\]
is then necessarily constant in time at $\r_0$ and under the matching equates to $m$ in the exterior. The data and free functions need to satisfy various open conditions to ensure that $m$, $V$, $\dot T$ and so on are positive but this is easily done.
\\\\
{\bf Acknowdgements}
\\\\
FM thanks Departamento de Matem\'atica, Instituto Superior T\'ecnico for hospitality, FCT (Portugal) for grant FRH/BPD/12137/2003 and CMAT, 
Universidade do Minho, for support. 
JN was partially supported by FCT (Portugal) through the Program POCI 2010/FEDER and by the Project POCI/MAT/58549/2004. 
PT thanks Departamento de Matem\'atica, Instituto Superior T\'ecnico, where this work was initiated, for hospitality.

\end{document}